\documentclass[fleqn,10pt]{wlscirep}
\usepackage{subfig}
\usepackage{url}
\usepackage{caption}
\usepackage{float}
\usepackage{subfloat}
\usepackage{cite}
\usepackage{tabularx}
\usepackage{tabulary}
\usepackage{rotating}

\title{
The evolution of k-shell in syndication networks reveals financial performance of venture capital institutions} 

\author[1,*]{Ruiqi Li}
\author[1]{Jing Liang}    
\author[2,*]{Cheng Cheng} 
\author[1]{Xiaoyan Zhang} %
\author[3]{Longfeng Zhao} 
\author[4]{Chen Zhao}   
\author[5,*]{H. Eugene Stanley} 

\affil[1]{UrbanNet Lab, College of Information Science and Technology, Beijing University of Chemical Technology, Beijing 100029, China}
\affil[2]{School of Finance and Statistics, Hunan University, Hunan 410082, China}
\affil[3]{School of Management, Northwestern Polytechnical University, Xi'an 710072, China}
\affil[4]{College of Information Technology, Hebei Normal University, Hebei 050024, China}
\affil[5]{Center for Polymer Studies and Physics Department, Boston University, Boston, MA 02215, USA}
\affil[*]{correspondings. lir@buct.edu.cn, ccheng@hnu.edu.cn, hes@bu.edu}    


\begin{abstract}
Venture capital (VC) is a relatively newly emergent industry that is still subject to large uncertainties in China. Therefore, building a robust social network with other VC institutions is a good way to share information, various resources, and benefit from skill and knowledge complementarity to against risks. Strong evidences indicate that better networked VC institutions are of a better financial performance, however, most of previous works overlook the evolution of VC institutions and only focus on some simple topology indicators of the static syndication network, which also neglects higher-order network structure and cannot give a comprehensive evaluation. 
In this paper, based on VC investment records in the Chinese market, we construct temporal syndication networks between VC institutions year by year. As k-shell decomposition considers higher-order connection patterns, we employ k-shell as an evaluation of the influence of VC institutions in syndication networks. By clustering time series of k-shell values, the VC institutions in China fall into five groups that are quite different from each other on financial performances and investment behaviors. This, in turn, proves the power of our method that only based on proper sequential network properties, we can reveal their financial investment performance. Compared to other network centrality measurements, k-shell is a better indicator that is indicated by a smaller intra-group distance and a larger inter-group distance. 
\end{abstract}
\begin{document}

\flushbottom
\maketitle

\thispagestyle{empty}

\section*{Introduction}
Since the origin of the venture capital (VC) industry after World War II, there already had been several boom-and-bust cycles in the Western world \cite{gompers2004venture}, and the institutionalization of the VC industry is commonly dated back to three events in 1978 and 1980 \cite{2007whom}. By contrast, the history of Chinese VC institutions, as well as state-owned ones, only dates back to 1985 \cite{lu2013venture,website:PEdata}. Venture capital is still a relatively newly emergent industry in China, in which government policies keep changing, the governance structure is immature, and information asymmetry always bothers investors \cite{luo2015guanxi,website:BankVC}. A few years ago, the Chinese State Council publicly calls for more government financing, especially commercial banks, in venture capital to get the state to take part in the nation's technology boom \cite{website:BankVC}.  All those factors mentioned before make the Chinese VC investment environment highly uncertain and short-term rational calculation of investors often in vain \cite{luo2016ERGM}. Therefore, building a robust social network with other VC institutions is a good way to access information \cite{bygrave1987syndicated,burt2009holes}, other VC institutions' deal flows on the reciprocal basis \cite{lerner1994syndication}, resources \cite{bygrave1988structure,ahuja2000collaboration,katila2008swimming} and benefit from the skill and knowledge complementarity \cite{uzzi1996sources,brander2002venture,yao2022effects}, diversity \cite{lerner1994syndication,podolny2001networks,manigart2002european} to against uncertainties \cite{bygrave1987syndicated}, free riding and opportunism behaviors \cite{williamson1985,2007whom,wilson1968theory,sorenson2001syndication,manigart2002european}, improve screening \cite{sah1986architecture}, and gain reputations \cite{smith2010venture,hochberg2010networking,sathe2010ventureSecrects}. 

\begin{figure}[ht] \centering
\includegraphics[width=\linewidth]{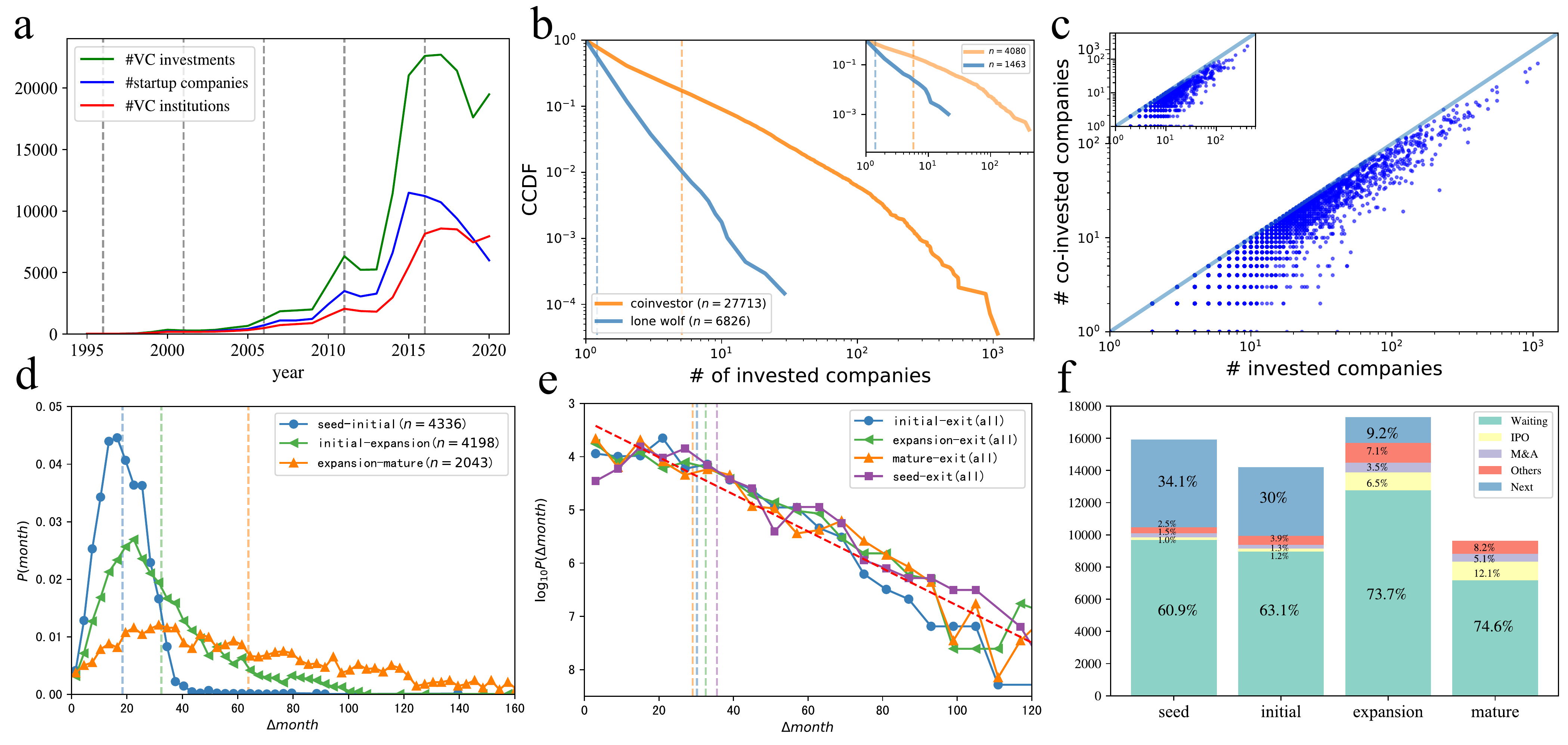} 
\caption{(a) The number of VC investments, the number of startup companies got funded, and the number of VC institutions in China. (b) The complementary cumulative distribution function (CCDF) of the number of invested startup companies by VC institutions that have (``co-investor'') and have not (``lone wolf'') ever made joint-investment with others, respectively. 
(c) The number of joint-investments versus the number of all investments of every co-investor. Each dot represents a VC institution. 
(d) The distribution of waiting duration of consecutive VC investments between different stages. In the legend, $n$ denotes the number of corresponding cases. (e) The distribution of duration between the final stage of VC investment and the exit, which includes all types with both successful (IPO or M\&A) and unsuccessful ones (e.g., liquidation, buybacks). (f) The fraction of startup companies that receive the next round of financing (denoted as ``Next''), exit through IPO or M\&A or other less successful ways (``Others''), and also the ones that has no further updates on their status (``Waiting''). }
\label{fig1}
\end{figure}

In China, influenced by the phenomenon of ``\textit{guanxi}'' \cite{luo1997guanxi,luo2019syndication} that values long-term social relationship and long-term financial returns, the networking behavior of Chinese VC institutions is quite ubiquitous and profound (see Fig. \ref{fig1}b,c). 
Chinese VC institutions tend to form ``clan-like group'' \cite{Hsu1963,Fei1992,luo2016ERGM}, which is different from western ``club-like group'' \cite{Hsu1963,boisotChild1996}, and applies different rules of social exchanges for different types of \textit{guanxi} \cite{Fei1992}. With the formation of this kind of communities \cite{shang2022local}, the order would emerge: there will be ``big brothers'' \cite{farhChen2000,luo2016ERGM} that are usually top-tier VC institutions in the core of the network with high reputation and great influence, and some ``rookies'' at the fringe of the network \cite{luo2016ERGM,gu2019exploring}. 
Big brothers also tend to have better deal flows, better accessibility to information \cite{2007whom}, and stronger bargaining power to the portfolio companies they invest in \cite{hsu2004entrepreneurs}. For example, some evidence showed that offers made by VC institutions with a high reputation were three times more likely to be accepted and at a 10-14\% discount \cite{hsu2004entrepreneurs}.
If a young VC institution can hop onto an eventually successful deal led by a marquee firm (i.e., big brother), it can gain some of the marquee firm's luster and become visible to the public \cite{smith2010venture,sathe2010ventureSecrects}. And sometimes, for new entrants, the existing densely connected syndication network (or better termed as cliques) formed by incumbent VC institutions can be a potential barrier for them to enter the market \cite{hochberg2010networking}. Generally, more densely networked market is harder to enter for a new entrant \cite{hochberg2010networking}. Establishing connections with incumbents to access their local information, expertise, or contacts is generally a good way to get over such barrier, yet the result still depends on the strategical reactions of incumbents by balancing between potential gains (e.g., access to the home market of the new entrant \cite{hochberg2010networking} or pooling capital for risky tryouts \cite{luo2016ERGM}) and pressures from its peers with the afraid of gradually loosing bargaining power due to competitions by introducing outsiders into their local market \cite{hochberg2010networking}. VC institutions tend to routinely cooperate with a small set of VC institutions, which also indicates that syndication relationships are relatively exclusive and stable \cite{2007whom,hochberg2010networking,luo2016ERGM}. 
So the position and the evolution of the position (i.e., another type of growth trajectory) of a VC institution in the syndication network is critical for its investment activities and consequently its financial performance \cite{2007whom,hochberg2010networking}. Thus it is crucial to identify the position and evaluate the influence of VC institutions in the syndication network with proper network indicators.

The syndication network of VC institutions has been studied for decades, however, most previous works tend to use some simple centrality indicators of the topology of a static network (such as degree, in-degree, out-degree, betweenness, etc.) as a proxy of the influence of a VC institution  \cite{lerner1994syndication,sorenson2001syndication,2007whom,dang2011venture,hochberg2010networking,jaaskelainen2014networks}. However, these indicators cannot give a very proper evaluation of the influence of the VC institutions, because not only whom you know \cite{2007whom} and how your friends connect with others, which can be better described by higher-order structure \cite{Makse2010kshell,jure2016higher}, rather than just how many people you know matters a lot. 
More importantly, as the VC syndication network is growing and evolving, static descriptions based on snapshots lost important information about VC institutions on how it evolves, and how it affects the investment performance over a longer period. 
And previous works usually construct a directed network between VC institutions, where, for each joint-investment, the lead investor (usually identified as the one with the largest investment amount) points to other co-investors (there are no links between those non-dominant co-investors) \cite{2007whom,dang2011venture,hochberg2010networking}. Constructing such directed networks neglects the reciprocity nature of the mutual cooperation relatioship and will lose information between other non-lead co-investors in the joint-investment. 
Besides, especially in the Chinese VC industry, sometimes the lead investor cannot be inferred from the investment amount \cite{luo2016ERGM,gu2019exploring}. When a case is too risky, the average amount invested is generally less \cite{bygrave1987syndicated}, and the lead investor may ask other followers to have a larger share; though aware of the potential risk, followers may still take the deal in exchange for establishing a better \textit{guanxi} with big brothers and potential long-term financial returns \cite{luo2016ERGM}. 
In addition, such financial related data may not always be available to researchers, especially for some fast growing markets in developing countries, which is also the case in other fields \cite{dong2016population}. 
By contrast, the syndication information (i.e., whether two VC institutions have joint-investment or not) is easier to obtain and less noisy. 

In this work, 
with the Simutong dataset that records venture capital investment events 
in the Chinese VC market \cite{website:PEdata}, we construct syndication networks between VC institutions year by year. We employ k-shell decomposition \cite{Makse2010kshell,PRL2006kcore,li2017effects} as an evaluation of the influence of VC institutions in syndication networks, and validate its effectiveness via comparing with other popular centrality measurements. From the evolution of their k-shell values, VC institutions in the Chinese market are classified into five groups that are quite different from each other on financial investment performances. This in turn proves the power of our method that only basing on proper sequential network properties, we can reveal their financial performance.  The results of this work would be helpful to give suggestions for limited partners on their investments in VC institutions, and for startup companies on approaching VC institutions.

\section*{Results}
\subsection*{A glance at VC industry in China.}
Since the origins of the modern private equity industry (VC is a type of private equity) in 1946, there have been several boom-and-bust cycles worldwide \cite{gompers2004venture}, while, in China, the VC industry is still a newly emergent industry with a much shorter history. 
In 1985, the first VC institution in China (the China New-tech Venture Capital Corporation, CNVCC) was established, which was fully owned by the Ministry of Science and Technology \cite{lu2013venture}. In 1992, IDG entered the Chinese VC market, which is the first foreign VC institution in China, via developing a good relationship with local governments \cite{lu2013venture}.  
At the very first, with strong information asymmetry and the lacking of relevant laws and regulations \cite{lu2013venture}, there were a few VC institutions (see Fig. \ref{fig1}a). 
The number of VC institutions that have investment activities in each year grows substantially after 2013 (see the red line in Fig.\ref{fig1}a). 
The number of VC investments and the number of start-up companies that got investments from VC institutions increased with a similar trend and roughly manifests five-year boom-and-bust cycles (see green and blue lines in Fig.~\ref{fig1}a; for clarity, in this work, we only use the term ``company'' to refer to the start-up enterprise that got invested by a VC institution). 
And since 2016, the number of VC institutions that made investments keeps at a high level, while the number of startup companies is decreasing, which may resemble the situation in a few years since 1985 in the United States when too much money was chasing too few deals \cite{bygrave1987syndicated}.  
With the rapid growth, it is urgent to gain a better understanding of the Chinese VC market. 

Networking between VC institutions via making joint-investments is ubiquitous and profound. 
In the Chinese VC market, the majority VC institutions are the ones ever made joint-investment with others (see Fig. \ref{fig1}b), there were only around one-fourth of VC institutions that made investments solely, which are referred to as ``lone wolf'', until 2013 (see inset in Fig.~\ref{fig1}b) and around one-fifth of them until 2020(see Fig. \ref{fig1}b), and the average number of investments made by them are quite small (see the blue dashed vertical lines in Fig.~\ref{fig1}b). 
The number of startup companies invested by ``lone wolves'' only takes up a relatively small fraction of the whole market (roughly 15\%), 
by comparison, the vast majority are made by the ``co-investor'' VC institutions. 
Furthermore, for ``co-investor'' VC institutions, we find that joint-investments take up a significant fraction (the average is around 67\%) of their investment activities, especially for big VC institutions (see Fig.~\ref{fig1}c). It is almost impossible to observe a big VC institution with a small number of joint-investments. And such a trend is also stable over time, the situation in 2013 (see the inset in Fig. \ref{fig1}c) is qualitatively the same with the case in 2020 (see Fig. \ref{fig1}c). 
Whether growing big requires extensive cooperation with others or extensive cooperation makes a VC institution grow big is worth closer future studies. 
In this work, we only focus on VC institutions that ever made joint-investments with others.

As VC investments involve long-term engagement and possible further capital investment, we also analyze the distribution of duration between consecutive investment stages \cite{bygrave1987syndicated,yao2022effects} (see Fig. \ref{fig1}d). After making investments in the seed stage of a startup, on average, it takes roughly 18.6 months for the startup to get to the initial stage. From the initial stage to the expansion stage, the average duration is 33.4 months, and 65.1 months to get to the mature stage. And the tail part of the distribution is wider for the latter two cases. 
However, if a startup receives no further investment, the distribution of duration to exit is quite similar and manifests a long tail, which can be well approximated by an exponential distribution $P\propto e^{-3.2\Delta month}$. This indicates that with a longer time, the probability of exit will drop rapidly (see Fig. \ref{fig1}e). 
More intriguingly, for both successful (IPO or M\&A) and unsuccessful (liquidation, buyback, etc.) types of exit, their distributions all can be well approximated by an exponential distribution (see Fig. \ref{fig.appendix.exit} in Appendix), whose coefficients are quite close to the ensemble one with all types of exit (see Fig. \ref{fig1}e). 
The fraction of startups going to the next stage or exit in different ways are shown in Fig. \ref{fig1}f. Roughly one-third of startups after receiving seed stage or initial stage investment can go to the next stage, and this probability drops to roughly one-tenth for startups in the expansion stage. 
The fraction of IPO and M\&A all increases for startups at the expansion or mature stages, meanwhile, the fraction of unsuccessful exits also increases. 
Yet, a cruel fact is that the majority of startups are the ones with no further updates on their status, which may receive no further investments or no longer exist -- roughly 60\% for seed and initial stages, and increases to more than 70\% for expansion and mature stages (see Fig. \ref{fig1}f). This also indicates that the risk of failure does not necessarily decrease for expansion and mature stages.  

\subsection*{Constructing temporal syndication networks.}
Until the end of 2013, there were more than 33,000 VC investment activities that span most industries and regions in China. 
Each investment record details the name of the investor (usually a VC institution, sometimes an individual investor or angel), the start-up company got invested, and basic information about the start-up, including its industry category and headquarters location, the date of investment, the investment amount (we converted the foreign currency into RMB), investment stage (which includes seed, initial, expansion, and mature). 
Since we only focus on VC institutions, we eliminate those investments made by individuals and undisclosed investors. After filtering such records, there are more than 30,700 investment records left, more than half of which are co-investments. 

\begin{figure}[ht] \centering
\includegraphics[width=0.7\textwidth]{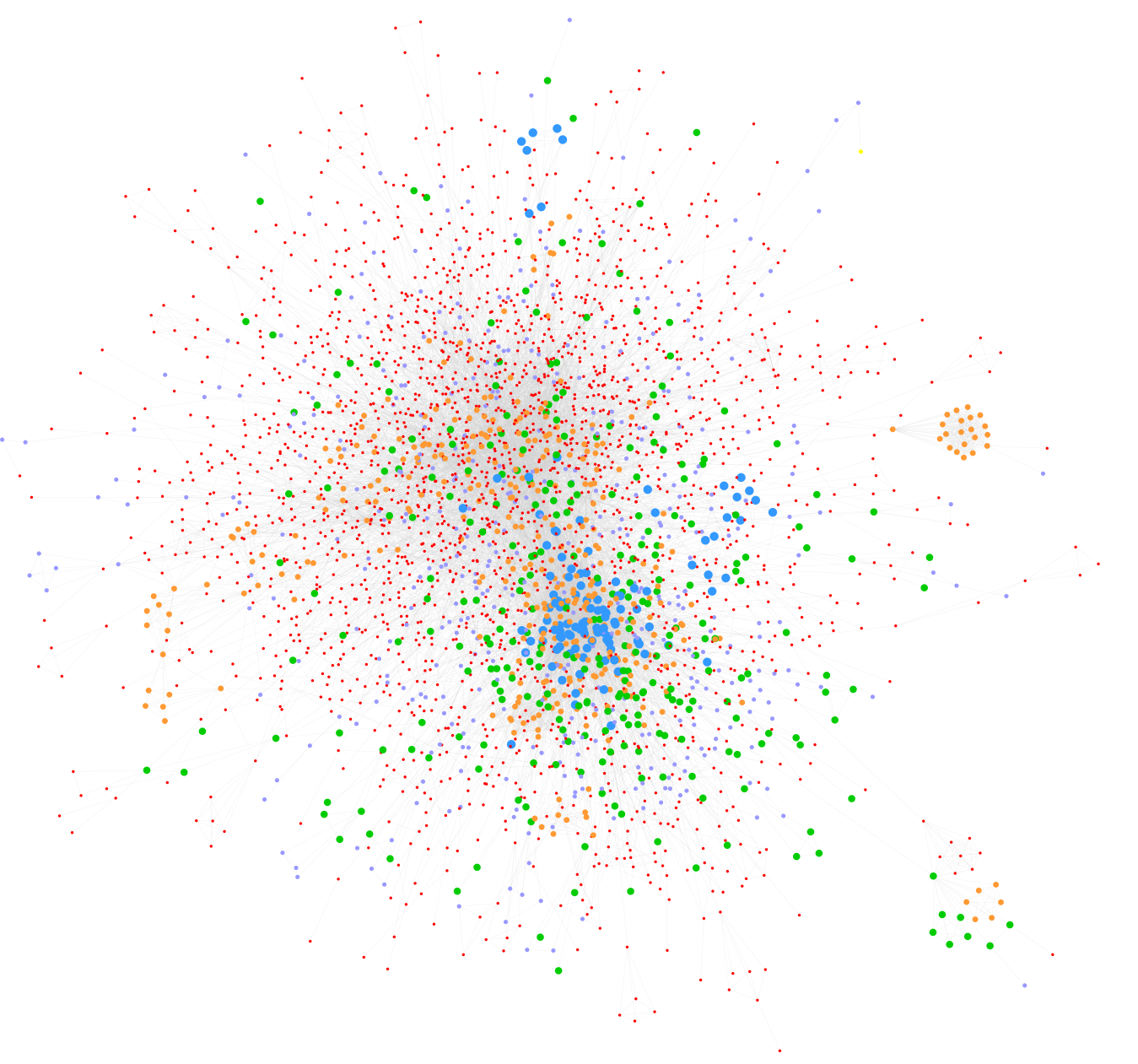} 
\caption{The syndication network in the year 2013. 
The color of nodes corresponds to the classifications obtained from clustering time series of their k-shell values over past years. Blue, orange, green, purple, and red corresponds to big brothers, rising stars, unsuccessful first movers, outsiders, and rookies, respectively. The size of nodes is also in line with classification labels.}
\label{figNetwork}
\end{figure}

Based on investment activity records, we identify joint-investment between VC institutions as the ones made on the same start-up company on the same date. 
In the syndication network, nodes represent VC institutions, and an edge between two VC institutions in the network indicates that they two had a joint-investment. 
We do not regard the investments at different rounds as joint-investment since the investments between different rounds are not necessarily direct collaboration and more uncertain to infer whether they two have social connections. For example, some VC institutions that make investments in earlier stages may exit in later rounds when new investors buy the shares of existing investors. Although all the VC institutions that made investments in the same company might have a position on the startup company's board of directors, the nature of the relations between them is also harder to identify. 
In each year, we construct an accumulative syndication network, i.e., if two VC institutions ever had a joint-investment in or before that year, then they two will have a link in the network. The reason for constructing an accumulative network is that a previous joint-investment and collaboration relationship might last for a much longer time \cite{beckman2004friends}, and a successful collaboration might further strengthen the relationship \cite{kaplan2005private,gu2019exploring}. Constructing a dynamic network that only comprises join-investment relation in a certain year will lose such long-term information, and setting a time window also can be arbitrary. To avoid further bias, we do not make such a setting. The interplay between the memory effect and the forgetting effect regarding collaboration relation is an important topic that is worth future investigation.    
In such a way, we get twenty-four networks from the year 1990 to 2013 (see Fig.~\ref{figNetwork}). 

Before 2014, there were around 5,543 VC institutions in China that have at least one investment record, and 4,080 of them have made at least one joint-investment with one another, with only 644 of which were not in the giant component of the network (see Fig.\ref{figNetwork}).
The reason why we focus on syndication networks until the end of 2013 is due to the fact that for better evaluating the investment performance, we need several more years of data on exit events (see Fig. \ref{fig1}e). In this study, we use exit events data before the year 2021, with seven more years, most of the startups that got invested before 2014 would exist. 

\subsection*{Evolution of k-shell of VC institutions.}
\begin{figure}[ht] \centering
\includegraphics[width=\textwidth]{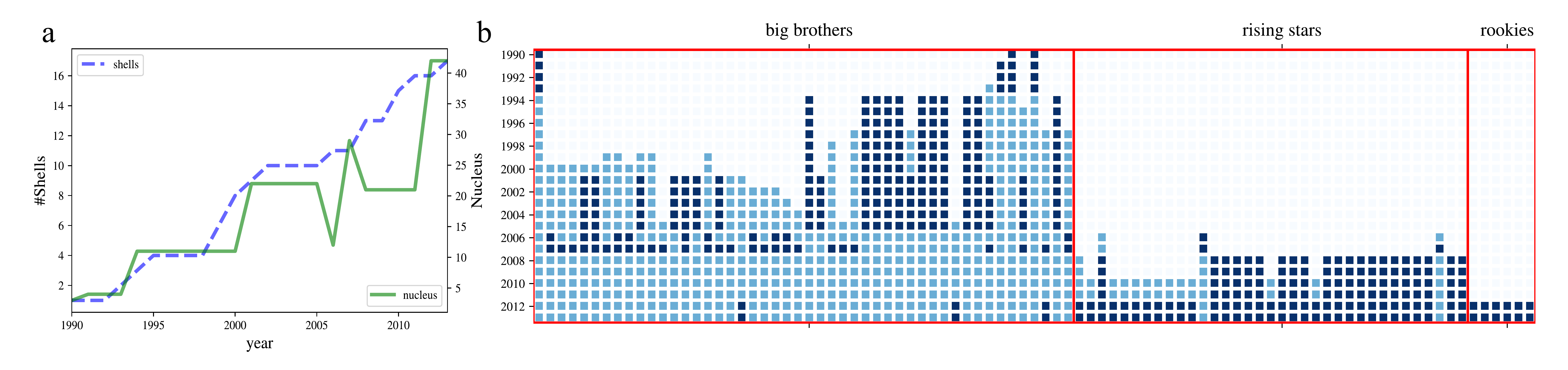}
\caption{(a) The evolution of the number of shells (blue dashed line that corresponds to the left y-axis) and the size of the nuclei (i.e., the most central core, green line that corresponds to the right y-axis) of syndication networks. (b) The VC institutions that ever entered the nuclei from 1990 to 2013. Each column corresponds to a VC institution, and each row corresponds to a certain year. The dark blue square indicates that the corresponding VC institution was in the nuclei, and the light blue square indicates entering the Chinese VC market but not in the nuclei that year.} 
\label{fig.appendix.kshell}
\end{figure}

Previous work \cite{2007whom} indicates that the centrality of nodes, i.e., the influence (or importance) of a node in a network, is related to their investment performances, yet, the growth trajectory of a VC institution is largely unknown. Most previous works use degree, betweenness, or eigenvector centrality as proxies of the prominence of VC institutions, and only focus on static networks at a snapshot. 
With recent development in network science, k-shell (also named k-core) has been proven to be a better indicator on measuring the influence of a node \cite{Makse2010kshell,PRL2006kcore}, and it naturally reveals a core-periphery structure (see Appendix for more details of k-shell decomposition). Generally, the nodes in the central core will form a complete graph, in which every node is connected with others \cite{Makse2010kshell,zhang2023hego}. 
We discover that along the increase of the whole network, the number of shells in the whole network, which represents a sort of hierarchical structure, becomes larger over past years (see the blue dashed line in Fig. \ref{fig.appendix.kshell}a). The number of nodes in the most central core (termed the nuclei) increases over time as well but has some fluctuations (see the green line in Fig. \ref{fig.appendix.kshell}a). 

By applying k-shell decomposition on all twenty-four networks from the year 1990 to 2013, for each VC institution, we can get the evolution of its k-shell value from the first year when it entered the Chinese market to the year 2013. Such a time series encodes great information regarding the growth trajectory of VC institutions in the syndication network, and it can be related to the investment performance of VC institutions \cite{2007whom}. 

\subsection*{Classification of VC institutions.} 
\begin{figure}[hbt!] \centering
\includegraphics[width=\textwidth]{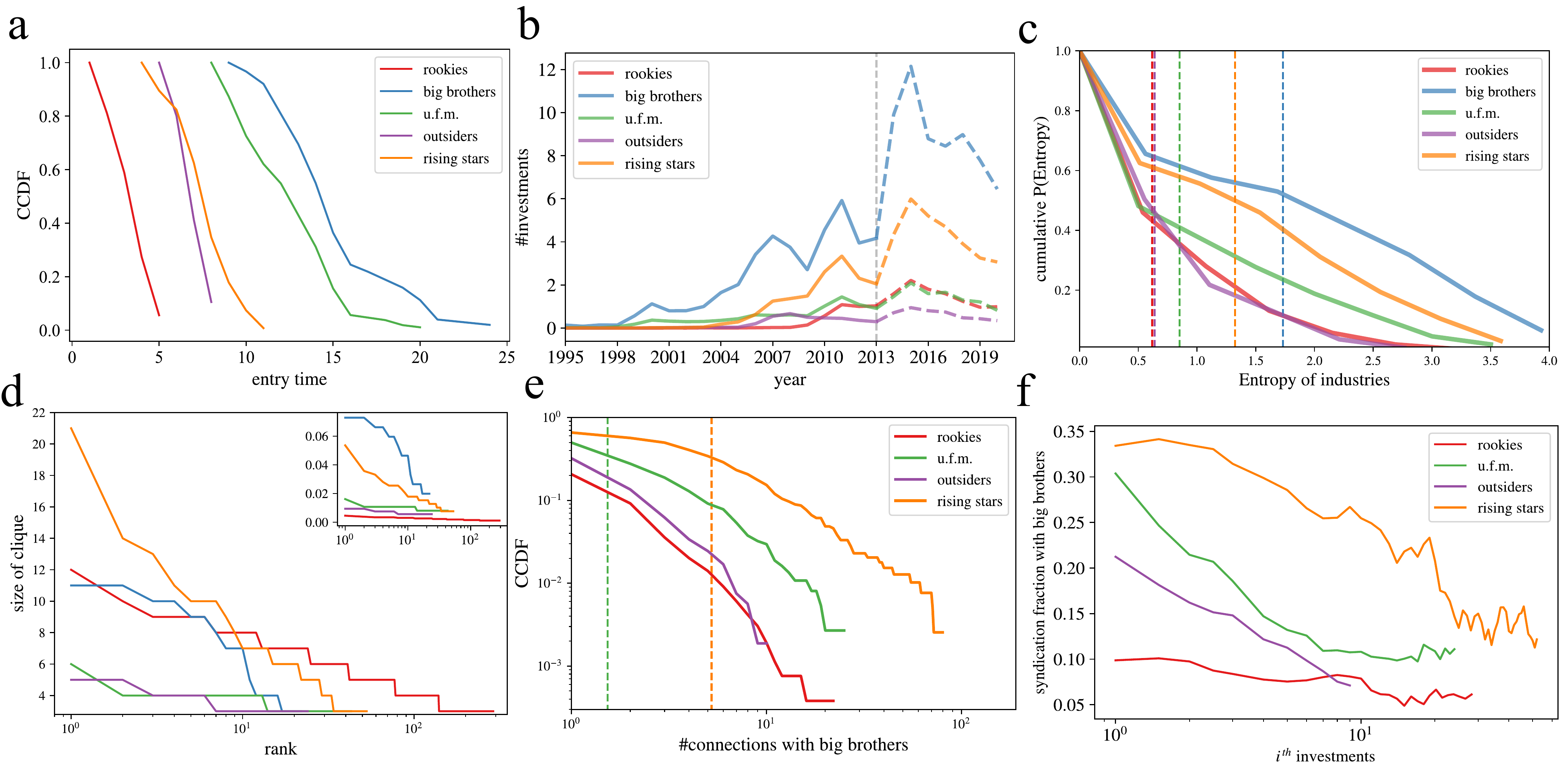}
\caption{(a) The distribution of the duration of practicing since their entry year. Big brothers and unsuccessful first-movers (u.f.m.) are of similar age; rising stars and outsiders are also of similar age. (b) The average number of investments, which includes both solo and joint-investments, made by VC institutions in each group. Data after 2013 is shown as dashed lines, which still have a clear separation between groups and further confirm the predictive power of the partition. In addition, there are two interesting bifurcation phenomena: big brothers and u.f.m. diverged after the year 1998; rising stars and outsiders further diverged after the year 2007. 
(c) The distribution of industry entropy of VC institutions in each group. Big brothers and rising stars invest in more types of industry and have a larger diversity. A similar trend holds for the region entropy (see Fig. \ref{fig.appendix.regionsdiversity} in the Appendix). 
(d) The distribution of cliques formed by VC institutions within each group, and these cliques are not overlapping with each other; (inset) the corresponding distribution that is normalized by the size of each group.  
(e) The cumulative probability distribution of the number of connections to big brothers of VC institutions in other groups. The average number is indicated by vertical dash lines. For outsiders and rookies, as their average value is smaller than 1, which is not shown in the figure. 
(f) The moving average of syndication fraction with big brothers drops as the times of investment increase. The time window for the moving average is five. 
}
\label{fig.groups}
\end{figure}

Due to the fact that the time-series of k-shell values are length variant (e.g., older VC institutions have a longer time series since their entry year, while younger firms have a shorter sequence), we extract four features of each sequence to embed time series in a lower dimensional vector space -- the length of the sequence ($years$), the k-shell value in the year 2013 ($ks_{2013}$), the difference between $ks_{2013}$ and the initial k-shell value in the entry year (defined as $\delta_{ks} = ks_{2013}-ks_{entry}$), and also the area below the curve of the time-series (i.e., $area=\sum_{j=entry}^{2013}ks_j$). In order to make the data of different measurements comparable in magnitude, we scale the data first (see Methods).
Then through the hierarchical clustering algorithm, we reveal that there are five groups of VC institutions in the Chinese VC industry (see Fig.~\ref{fig.groups}). 
And surprisingly, although we only use proper network indicators to make the classification and not using financial-related data at all, these groups have quite different investment performances (see Table~\ref{table.Group}).

\begin{table*}[!htbp] \footnotesize  \centering
\caption{Properties of VC institutions in different groups.}
\label{table.Group}
\begin{tabulary}{1.6\textwidth}{c|cccc|cccccc|cc} 
\hline
\hline  
Description &$N$ & $\langle Years\rangle$ & $\langle ks_{2013}\rangle $ & $\langle \delta_{ks}\rangle $ &$\langle Amt_{tot}\rangle$ &  $\langle Amt_{IPO}\rangle $ & $\langle R_{tot}\rangle$ &  $\langle R_{IPO}\rangle $ &  $\langle \#C_{IPO}\rangle $ & $\langle \#C_{tot}\rangle$  & $\langle IPO\rangle$ & $\langle HEI\rangle$\\ 
\hline
big brothers & 151  & 14.311 & 11.033 & 6.702 & 4442.503 & 1365.455 & 41.417 & 9.093 & 5.483 & 23.974 & 0.260 & 0.945 \\
u.f.m.       & 372  & 11.828 & 3.446  & 1.543 & 875.065  & 256.443  & 9.266  & 1.755 & 1.234 & 6.694  & 0.185 & 0.596 \\
rookies      & 2633 & 2.730  & 3.513  & 0.659 & 245.627  & 57.761   & 3.933  & 0.732 & 0.607 & 2.090  & 0.239 & 0.466 \\
outsiders    & 532  & 6.318  & 2.395  & 0.682 & 391.468  & 43.112   & 3.671  & 0.774 & 0.615 & 2.667  & 0.228 & 0.469 \\
rising stars & 392  & 6.952  & 9.390  & 3.684 & 1714.302 & 429.966  & 15.656 & 3.332 & 2.393 & 9.954  & 0.258 & 0.706 \\ \hline
\hline  
\end{tabulary}
\caption*{Notes: $N$ is the size of the group; $\langle \cdot \rangle$ indicates the average value of entities in each group, for example, $\langle years\rangle$ is, on average, the number of years of VC institutions in the group since their entry of Chinese VC market; $\langle ks_{2013}\rangle$ is the final average k-shell value of VC institutions in 2013; $\langle \delta_{ks} \rangle$ is the difference between $ks_{2013}$ and k-shell value $ks_{entry}$ in the entry year, which is related to their growth trajectories; 
$\langle Amt_{tot}\rangle$ is the average total investment amount; $\langle Amt_{IPO}\rangle$ is the average investment amount for investments on startups that got listed; 
$\langle\#R_{tot}\rangle$ is the average number of rounds of investments (for a certain company, VC institutions may invest for several rounds); $\langle \#R_{IPO} \rangle$ is the average number of rounds of investments in the company which got listed; 
$\langle \#C_{tot}\rangle$ is the average number of companies they had invested (i.e., the size of their portfolio); $\langle \#C_{IPO}\rangle $ is the number of companies they had invested that got listed;
$\langle IPO\rangle$ is the average IPO rate of VC institutions, which equals the number of companies got listed $\#C_{IPO}$ divided by the number of all startups they ever invested $\#C_{tot}$; $HEI$ is the hawk-eye index proposed by us, which is defined in Eq.~\ref{eq.HEI}, and $\langle HEI\rangle$ the average of $HEI$ of VC institutions in each group.}
\end{table*}

Ideally, to measure financial investment performance, we should use return on investments (ROI) for each deal, or at least the internal rate of return (IRR) of each fund. However, such data are usually not accessible to researchers as VC funds generally only disclose it to their limited partners who invested in the fund. 
Some evidence shows that the average VC fund writes off 75.3\% of its investments \cite{ljungqvist2009scaling}, which implies that VC funds earn their capital gains from quite a small subset of their portfolio companies, especially those a few successful exit events via an IPO \cite{2007whom}.
So we employ mostly commonly used measurements of investment performance and proxy of VC institution's expertise -- the total number of startups in its portfolio (i.e., how many companies it ever invested, denoted as $\langle \#C_{tot}\rangle$ in Table~\ref{table.Group}), IPO rate (the fraction of companies got listed in VC institution's portfolio, denoted as $\langle ipo\rangle$), the total amount of capital invested by the VC institution ($\langle Amt_{tot}\rangle$), the total number of investment rounds ($\langle \#R_{tot}\rangle$) \cite{dimov2007requisite,sorensen2007smart,2007whom,dang2011venture,gompers2000money,brander2002venture,nahata2008venture}. 
And we added three more similar IPO-related indicators as measurements of investment performance -- the total amount of capital invested by the VC institution on the companies that got listed ($\langle Amt_{IPO}\rangle$), the total number of companies in its portfolio that got listed ($\langle \#C_{IPO}\rangle$), the total number of investment rounds on the companies that got listed ($\langle \#R_{IPO}\rangle$). Generally, with more investments in a startup that got listed eventually, the VC institution would gain a more favorable return. So, we also introduce a new measurement named ``Hawk-Eye Index'' (HEI) based on these variables to depict their foresight on investment, which is defined as 
\begin{equation}
 HEI=\frac{Amt_{IPO}/Amt_{tot}}{\#R_{IPO}/\#R_{tot}}.
\label{eq.HEI}
\end{equation}
HEI is proportional to the ratio of the amount invested in the companies got listed to the total investment amount, and inversely proportional to the ratio of the number of rounds of investment in the companies that got listed to the total rounds, which equals the total number of investments. It is worth noting that a VC institution might invest several rounds in the same startup company, which is also a reflection of the confidence and judgment of a VC institution on the future development of the startup. 
If HEI is high, it indicates that the VC institution invested more money in companies from a larger set of choices. A higher HEI indicates that the VC institution invests its capital more wisely and has better judgment in foreseeing the prospects of the startup company.

As shown in Table~\ref{table.Group},  
one group of VC institutions is the ``Big Brothers'' that invest a lot startups and invest with large amounts of capital. They entered the Chinese market relatively early, but they are not necessarily the oldest VC institutions in China (the ``unsuccessful first-movers'' denoted as u.f.m in Table \ref{table.Group} have a comparable entry time, also see Fig. \ref{fig.groups}a). Apart from having a highest IPO rate,  these big brothers also have the highest HEI, which indicates that they invest a significant fraction of their capital in a few startups, which got listed, out of a large set of choices. And they also have the highest investment diversity on both industry (see Fig. \ref{fig.groups}c) and regions (see Fig. \ref{fig.appendix.regionsdiversity} in Appendix). 

Another group is ``unsuccessful first-movers'' (u.f.m.), which are  also pioneers in the Chinese VC industry with an early entry year among all five groups (see Fig. \ref{fig.groups}a), yet their centrality is quite low in the syndication network, and the investment performances are not good (see Table \ref{table.Group}). Their average IPO rate is the lowest among all five groups, and their average HEI is not very high. 
We plot the average number of investments of each group every year since 1995 (see Fig.~\ref{fig.groups}b), it is clear that u.f.m. remains at a low level of investment activities. 
We can clearly see that before 1998, the u.f.m. was roughly at a similar level with big brothers; yet after the year 1998, these two groups diverged -- big brothers go up and u.f.m. keeps low, which is just similar to new entrants even in recent years (see the red line in Fig. \ref{fig.groups}b). 
Our results indicate that older VC institutions are not necessarily successful, although they were indeed able to raise successive funds, which is also the case in the Western VC industry \cite{sathe2010ventureSecrects}.
It is worth noting that we only use data before 2014 to make classifications, the dashed lines in Fig. \ref{fig.groups}b is a further testing of our results with recent data.  

Another group is ``rookies'' that are new entrants at the fringe of the network and invest in a few startups with the smallest amount and the lowest HEI among all five groups. 
Rookies take up the majority among all VC institutions, and they are also the youngest (see Fig. \ref{fig.groups}a) and may face fiercer competitions with peers. They tend to form more cliques (i.e., complete subgraph, in which nodes are connected to all other nodes) between themselves (see Fig. \ref{fig.groups}c). 


Another group is ``rising stars'', most of them are not pioneers like those big brothers or u.f.m., but they have the second-highest IPO rate, HEI, investment amount, and size of the portfolio. Compared to the big brothers, which have a setback during the financial crisis around the year 2008 (see blue line in Fig. \ref{fig.groups}b), rising stars seemingly spotted some opportunities and gained a better development momentum since then. And they tend to form many quite large cliques with each other (see Fig. \ref{fig.groups}c), which may indicate large-scale close collaboration and better information flow between them. Even when normalized by the size of the group (see inset in Fig. \ref{fig.groups}c), the size of their federation is still quite high, which is only slightly lower than those big brothers. Rising stars also have a pretty high diversity on invested industry (see Fig. \ref{fig.groups}c) and regions (see Fig. \ref{fig.appendix.regionsdiversity} in the Appendix). 

Apart from being most influential entities in the network \cite{Makse2010kshell}, recent advances also show that nodes in the most central core (i.e., with the maximum k-shell value) is critical for maintaining cooperative system \cite{wu2023rigorous} from collapse. 
The VC institutions that ever entered the nuclei (i.e., the most central core of the syndication network) are generally big brothers or rising stars. And it is worth noting that since 2008, the VC institutions in the nuclei are some rising stars, which also reflects the uprising trend of new forces and profound changing in the Chinese VC market. In addition, a few rookies also enter the nuclei via making broad connections with important players, and these rookies would highly probably become key players in the future as well.   

The last group is ``outsiders'', they have a similar entry year with rising stars, yet from the year 2007, differences between these two groups become larger and larger on the number of investments (see Fig.~\ref{fig.groups}b). Almost half of them do not have any connections with big brothers (see Fig.~\ref{fig.groups}d), and the average number of connections with big brothers is only 0.61, which is similar to rookies whose average number is 0.39. These VC institutions have the second-lowest average IPO rate and the second-smallest portfolio size, and after the year 2013, they continue to have a quite low level of investments (see the dashed purple line in Fig.~\ref{fig.groups}a).

Table~\ref{table.Group} and Fig.~\ref{fig.groups}e also indicate that the connections to big brothers might be a critical factor in their success. Rising stars have the highest number of connections with big brothers, and then u.f.m., which might be due to the fact that these two groups are of a similar age, while outsiders and rookies generally have a few connections with big brothers. Moreover, with the increase of the number of investments, the syndication probability with big brothers is decreasing, which indicates that with more experience accumulated, they might less rely on big brothers. 
Although previous works show that better-networked VC institutions experience significantly better investment performance \cite{2007whom,dang2011venture}, our work may serve as preliminaries to a possible theory towards the success or growth dynamics of small VC institutions.

In addition, from Fig.~\ref{fig.groups}b, we can clearly see two peaks in 2007 and 2011 which coincides with the hot secondary stock market and rapid appreciation of RMB in 2007 and economic bubbles in 2011 in China. As shown in Fig.~\ref{fig.groups}b, we can observe that after the year 2011, both big brothers and rising stars declined in the number of investments.
And almost all these groups hop on the opportunity in 2015, which is attributed by government support (e.g., ``Internet Plus'' and ``Made in China 2025'' initiatives), tech sector growth, and rapid economic growth.

\begin{figure}[hbt!] \centering
\includegraphics[width=0.5\textwidth]{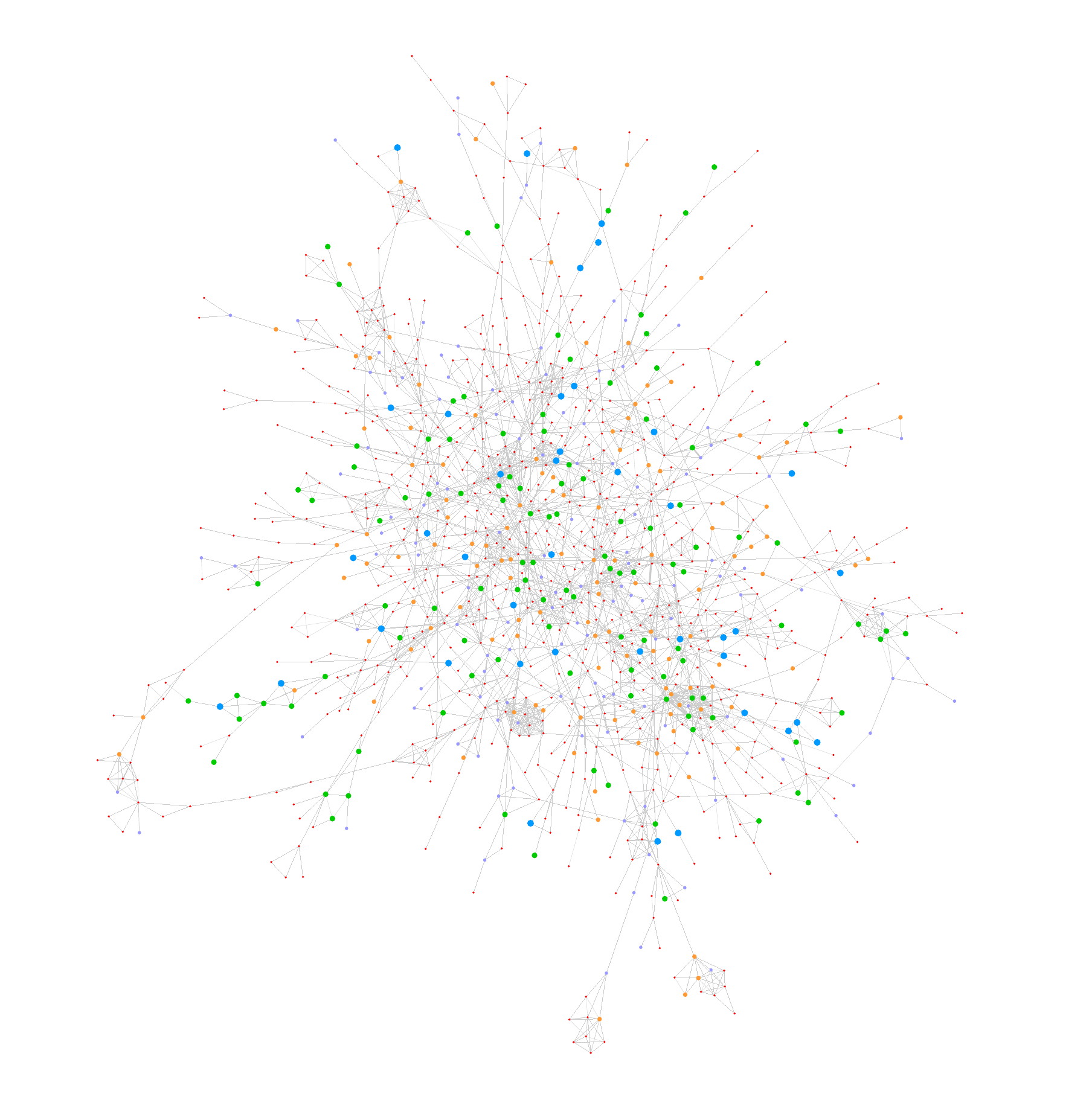}
\caption{
The common-shareholder networks between VC institutions. The color code and size of nodes are consistent with the ones in Fig. \ref{figNetwork}, which corresponds to different groups. The VC institutions that have no common shareholder with others are not shown in this figure. 
}
\label{fig.shareholder}
\end{figure}

Apart from making joint-investment, which are more public, having common shareholders is another type of important connection between VC institutions \cite{bygrave1987syndicated}. We construct a common-shareholder network via making queries on the board members of VC institutions (\url{https://www.qcc.com/}), and connect two VC institutions if they have at least one common shareholder (see Fig. \ref{fig.shareholder}). 

\begin{figure}[hbt!] \centering
\includegraphics[width=\textwidth]{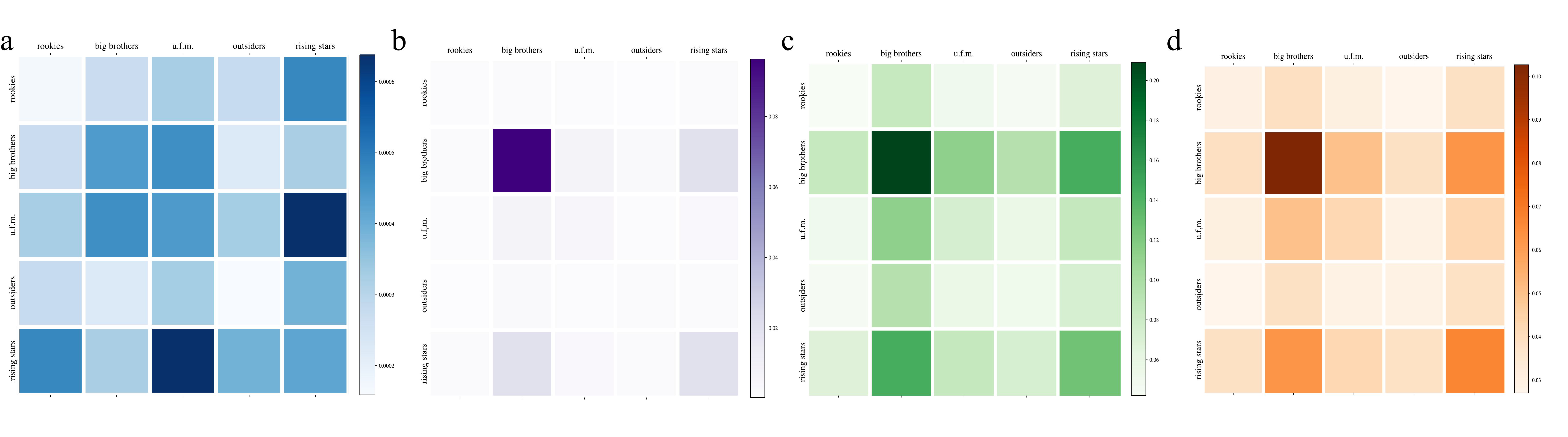}
\caption{The edge density on (a) common-shareholder relation and (b) joint-investment relation between VC institutions from different groups. 
(c) The industry similarity matrix between different groups. The industry similarity between VC $A$ and $B$ is calculated by the cosine similarity $S_{AB}=\sum_i^m A_iB_i/(\sqrt{\sum_i^m A_i^2}\sqrt{\sum_i^m B_i^2})$, where $m$ is the total number of industry types, and $A_i$ the fraction of investment of VC $A$ on the industry $i$, $B_i$ is the counterpart for VC $B$. Similarly, (d) region similarity can be defined, where $m$ is the total number of cities. Each entry in (c) and (d) represents the group average, i.e., averaging all VC pairs between two groups.}
\label{fig.edgedensity-similarity}
\end{figure}

Then we calculate the edge density 
between VC institutions from different groups (see Fig. \ref{fig.edgedensity-similarity}a), which equals the number of edges between VC institutions from two different groups divided by the product of the size of the two groups. For example, for the edge density between big brothers and rising stars 
equals the number of edges that connect a big brother on one end and a rising star on the other end divided by the product of the number of big brothers and the number of rising stars. 
It is intuitive to assume that shareholders of big VC institutions might be of a higher probability on the board of other VC institutions, however, the empirical results indicate that shareholders of big brothers only have some connections with u.f.m. and then big brothers, and quiet few connections with other types of VC institutions. Since big brothers and u.f.m. have a similar entry year,  it is not that surprising that they have more common shareholders. 
The most dominant one is the connection between u.f.m. and rising stars, which indicates that the board members of u.f.m. might just abandon previous unsuccessful institutions and focus on new VC institutions. While shareholders of rising stars also have a relatively high probability of being on the board of a rookie institution, which might also reflect their tendency of future strategies and bet on some new opportunities. Noting that the size of rookies is quite large, having a relatively high edge density between rising stars and rookies is nontrivial.  
And outsiders generally have a lower edge density with other institutions.

By contrast, the edge densities on joint-investment relation between groups (see Fig. \ref{fig.edgedensity-similarity}b) are quite different from the common-shareholders relationship (see Fig. \ref{fig.edgedensity-similarity}a). For syndication, connections between big brothers are of the highest density, 
and then the connections between rising stars, and between rising stars and big brothers are also relatively high. Such a discovery also indicates a rich club or elite club phenomenon \cite{gu2019exploring}. In addition, with all investments of a VC institution, we can get its industry profile, i.e., a vector details the fraction of investments of the VC institution in each industry. Then we calculate the industry similarity between any two VC institutions via the cosine similarity $S_{AB}=\sum_i^m A_iB_i/(\sqrt{\sum_i^m A_i^2}\sqrt{\sum_i^m B_i^2})$, where $m$ is the total number of industry classification, and $A_i$ the fraction of investment of VC $A$ on the industry $i$, $B_i$ is the counterpart for VC $B$. We can find that similarity patterns are highly correlated with syndication patterns but not common-shareholder patterns (see Fig. \ref{fig.edgedensity-similarity}a-c). The region similarity is similar to industry similarity but with smaller values (see Fig. \ref{fig.edgedensity-similarity}d). Results in Fig. \ref{fig.edgedensity-similarity}c-d shows that successful VC institutions are more alike to each other. 

\begin{figure}[ht] \centering
\includegraphics[width=0.4\textwidth]{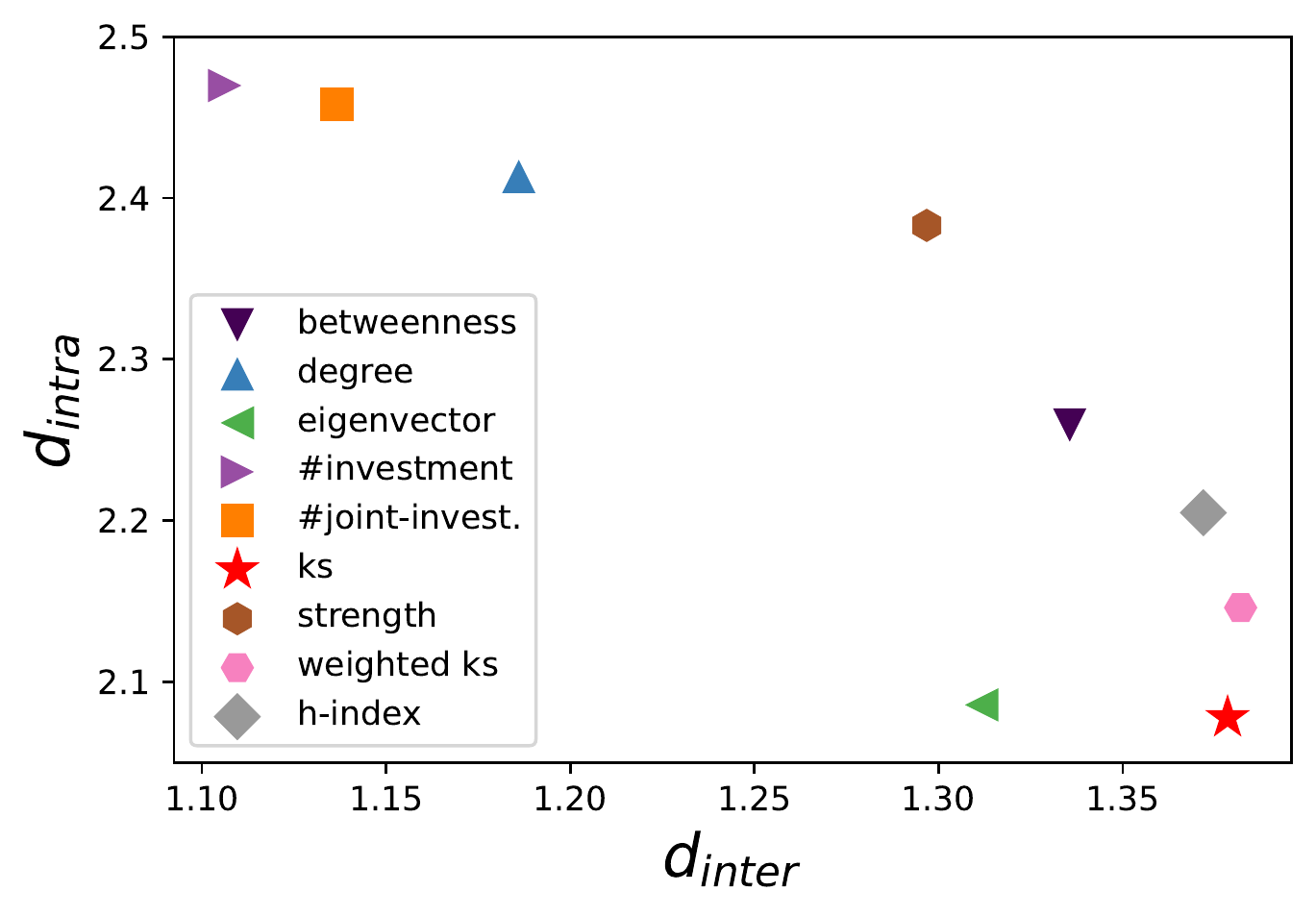}
\caption{The inter-group distance $d_{inter}$ and intra-group distance $d_{intra}$ for clustering results based on the evolution of different centrality measurements. The distance is measured on the metric space spanned by the $\langle IPO\rangle$ and $\langle HEI \rangle$, which are ensemble indicators derived from commonly used indicators in Table \ref{table.Group}. A good classification should satisfy greater differences between groups (measured as inter-distance $d_{inter}$ between the center of each group) and smaller distances within each group (measured as intra-distance $d_{intra}$ between the nodes in each group).}
\label{fig4}
\end{figure}

At last, after comparing the clustering results based on the evolution of other centrality measurements, including degree, nodal strength (i.e., the sum of weights of all links attached to a node), eigenvector centrality, betweenness, $h$-index \cite{lu2016h}, weighted k-shell \cite{garas2012k}, the number of investments, and the number of joint-investments, we discover that k-shell is the most suitable centrality indicator, which corresponds to a larger inter-group distance and a smaller intra-group distance than other centrality indicators (see Fig.\ref{fig4}). 

\section*{Conclusions and Discussions}
In this paper, based on the VC investment records from 1990 to 2013, we construct twenty-four syndication networks between VC institutions year by year in the Chinese market. We show that based on the evolution of k-shell values of VC institutions, we can classify VC institutions into five groups. These five groups are quite different from each other on investment performance, which in turn proves the power of our method -- only based on proper time series of topological features of VC institutions in syndication networks, we can reveal their financial investment performance. The results can provide references to limited partners when they decide which fund should they invest in. And we also show that the classification performance based on the evolution of k-shell is better than other centrality measurements, including degree, node strength, betweenness, $h$-index, and eigenvector centrality. 
Our results may also serve as preliminaries to a possible theory about the success or growth dynamics of small VC institutions, such as what are the impacts of connecting with a big brother and how the small company evolves with the interactions with big brothers.

And we also find that there are some VC institutions that only make investments at specific stages (e.g., mature or seed stage), 
which is another type of specialization and affected by syndication patterns \cite{yao2022effects} of VC institutions. In comparison, previous literature only focuses on industry specialization \cite{hochberg2015competition}. Syndication with a new partner would bring new information, expertise, and deal flow, with the classification of VC institutions presented in this work, more detailed dynamics can be analyzed, for example, it is worth investigating whether a rookie VC institution has a stronger tendency to syndicate with larger institutions when entering new stages or new industry, and whether such a choice will affect their future development. 
A successful collaboration might further strengthen the relationship and lead to new joint-investments \cite{kaplan2005private,gu2019exploring}, but a relationship generally will not last forever, which is hinted by the discovery in Fig. \ref{fig.groups}f, the interplay between the memory effect and the forgetting effect is an important topic that is worth future investigation.  In addition, we only focus on the cooperation relationship between VC institutions, yet there can also be competition relation between VC institutions \cite{bygrave1987syndicated,hochberg2015competition}, which can be modeled by a multi-relational network \cite{PNASmulti} as the nature of the links strongly affects the interaction dynamics on the networks \cite{li2013epidemic} and then might strongly affect their investment performance. There is also evidence showing that VC investments have impacts on urbanization \cite{bygrave1987syndicated,muller1977regional,li2021assessing}, the inflow of capital and concentration of high-tech companies are important factors driving the urbanization process (e.g., Silicon Valley and Route 128 \cite{dorfman1983route} in the United States) and should be incorporated in urban evolutionary models \cite{li2017simple}.

\section*{Methods}
\subsection*{Data}
We get access to detailed investment and exit records from the SiMuTong dataset \cite{website:PEdata}. The dataset we purchased is one of the most authoritative VC industry datasets in China. Each record in the investment dataset details the name of the investor (usually a VC institution, sometimes an individual investor or angel), the startup company that got invested, and the company's basic information, which includes industry and location, the date of investment, amount (we converted the foreign currency into RMB), share, stage. If two or more VC institutions jointly invest in the same startup company, this investment event will be shown as two or more records with the same date and other basic company information. 
Until 2013, there were more than 33,000 investment records, 862 of which are by individual investors, and around 1,400 records are with investor undisclosed. Since we only focus on VC institutions, we eliminate these investment records made by individuals and undisclosed investors. 


In the datset, there are some records whose company name is unrevealed, and we checked on-line manually to see whether they are revealed now, if so we update it into the dataset. 
For all those remaining records with the company name unrevealed, we use (industry, location, investment stage, and investment date) to identify a company, which is of a quite low probability that two different startup companies would the have same data on these four features. Thus even though we do not know the company name, we can still construct the syndication network correctly. 


The exit dataset contains records of IPO (Initial Public Offering), M\&A (Mergers and Acquisitions), equity transfer, buyback, liquidate, discharge of claims, among which, IPO is the best exit way with highest return, attention, and reputation \cite{smith2010venture}, and then M\&A; all other ways ot exit are less successful. 


\subsection*{Scaling of the data}
Scaling of the data (also termed as Standardization or Z-score Normalization) is a commonly used technique in machine learning when dealing with multiple variables that are with different magnitudes compared to each other. In our case, regarding different centrality variables, the magnitudes can vary from ten to thousands. For the data of each variable, we first calculate the mean and standard deviation of the entire vector, then subtract the mean and divided by the standard deviation, which can be expressed as  
\begin{equation}
\mathbf{v_{scaled}} = \frac{\mathbf{v}-\langle \mathbf{v} \rangle}{\sigma_{\mathbf{v}}},
\end{equation}  
where $\langle \mathbf{v} \rangle$ is the average value of the vector, and $\sigma_{\mathbf{v}}$ is the standard deviation. 
This process can be understood in an intuitive way as it is not changing the shape of the data but rather changing the scale of it.

\subsection*{Centrality measurements}
\subsubsection*{degree}
The degree of node $i$ \cite{anderson1992infectious} can be defined as $k_i=\sum_j A_{ij}$ where $A$ is the adjacency matrix of the network. The degree represents the number of neighbors a node has, which reflects the direct influence of this node on others.
\subsubsection*{eigenvector centrality}
The eigenvector centrality can be regarded as an extension of the simple degree centrality, where the neighbors of a node are not equally treated  \cite{newman2010networks}. It's formulated as 
\begin{equation}
v_i=\kappa_1^{-1}\sum_jA_{ij}v_j,
\end{equation}
where $\kappa_1$ is the largest eigenvalue of the adjacency matrix $A$. Thus $v_i$ can be large either because node $i$ has many neighbors or because it has important neighbors (or both).

\subsubsection*{node strength}
It is also referred to as node weight which is defined as the sum of weights attached to the ties belonging to a node, i.e., $s_i=\sum_{j}W_{ij}$ where $W_{ij}$ is the weight of the link between $i$ and $j$.

\subsubsection*{betweenness}
Betweenness \cite{freeman1977set} measures the extent to which a node lies on paths between other nodes which is important in social and spatial networks.
It's defined as 
\begin{equation}
b_i=\sum_{s,t} n_{st}^i/n_{st},
\end{equation}
where $n_{st}^i$ is the number of paths from node $s$ to $t$ that pass through $i$, $n_{st}$ is the total number of paths from node $s$ to $t$. \cite{newman2010networks}. 

\subsubsection*{Eigenvector centrality}
Eigenvector centrality is a classical measure of the influence of a node in a network. It assumes that the influence of a node $x_i$ is the average of its neighbors, which is defined as
\begin{equation}
x_{i}=\frac{1}{\lambda}\sum_{j\in V(i)}x_{j}=\frac {1}{\lambda}\sum_j^N A_{ij}x_{j}, 
\end{equation}
where $V(i)$ is the set of neighbors of node $i$, and $N$ is the total number of nodes in the network, and $A_{ij}=1$ if node $i$ and $j$ is connected, otherwise, $A_{ij}=0$. Such an equation can be rewritten in the matrix form that is the same with the eigenvector equation $\mathbf{A}\mathbf{x}=\lambda \mathbf{x}$. According to the Perron–Frobenius theorem, only the eigenvector associated with the largest eigenvalue are all non-negative, which meets the requirement of defining influence of nodes. 
A high eigenvector centrality means that a node is connected to many nodes who themselves also have a high influence. 

\subsubsection*{$h$-index}
$h$-index was first proposed for evaluating the academic influence of scholars based on the citation patterns of his/her publications \cite{hirsch2005index}, and it equals the maximum value of $h$ such that the given researcher has published at least $h$ papers that have each been cited at least $h$ times (i.e., there is no $h+1$ papers that each have a citation no less than $h+1$). 
Quite recently, the $h$-index got smartly adapted and applied to measure the influence of nodes in complex networks via replacing the number of publications with the number of neighbors and replacing the citation of each publication with the degree of each neighbor \cite{lu2016h}. The calculation of $h$-index can be defined as an operator, and iterating such an operator on the series of $h$-index of neighbors will converge to k-shell value \cite{lu2016h}. 

\subsubsection*{$k$-shell decomposition}
k-shell (also called k-core) decomposition starts by iteratively removing all nodes with minimum degree $k_{min}$ until there is no node left with $k\leq k_{min}$ in the network. The removed nodes are assigned with $ks=k_{min}$ and are considered in the first layer/shell, in most cases, $k_{min}=1$. In a similar way, nodes with current minimum degree $k_{min}+1$ are iteratively removed and assigned with $ks=k_{min}+1$. This decomposition process continues removing higher shells until ending up with a core all above a certain remaining degree (i.e., the central core, which will be removed at last). Isolated nodes are usually assigned with $ks=0$, and we omitted such nodes in this work. k-shell method decomposes the network into ordered shells from the core to the periphery, researchers found that nodes in the central core of the network are more influential than the periphery nodes \cite{Makse2010kshell}.

\begin{figure}[hbt!] \centering
\includegraphics[width=0.8\textwidth]{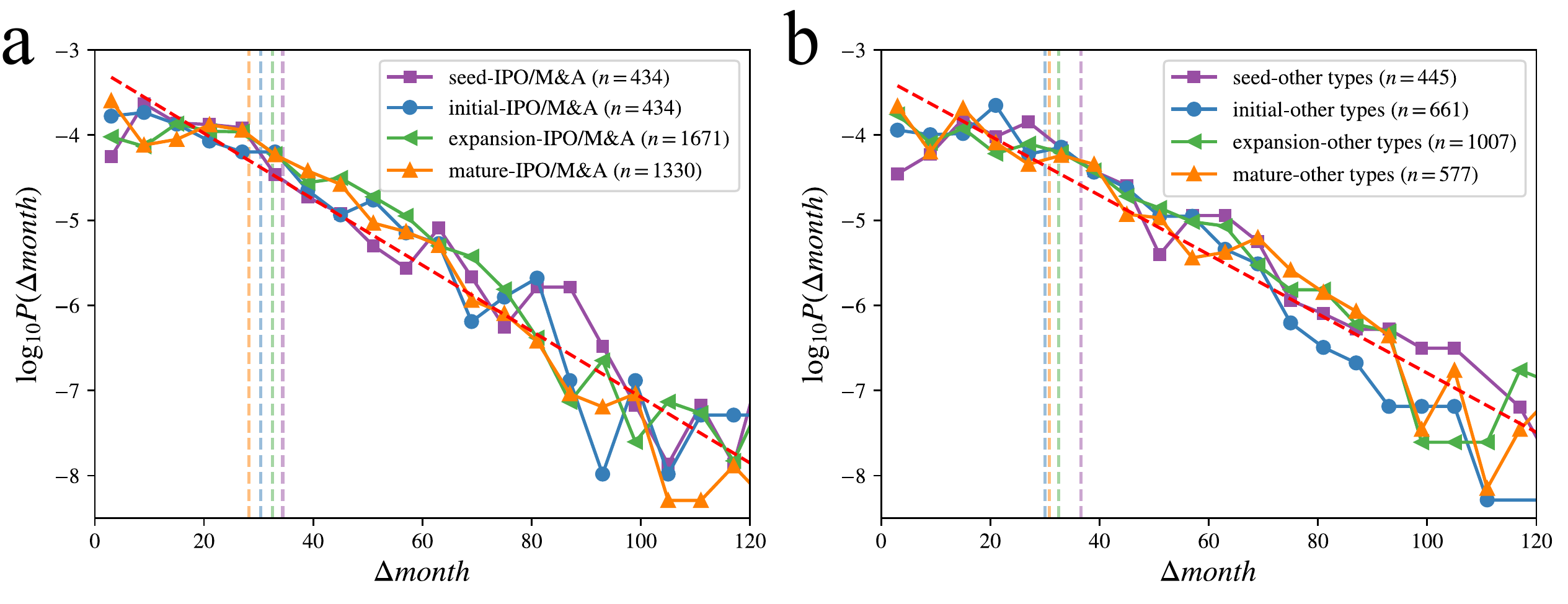}
\caption{The distribution of duration between a previous stage to (a) successful types of exit, i.e., IPO or M\&A, and (b) unsuccessful types of exit, including liquidation, and buyback. The distribution can all be well approximated by an exponential function $P\propto e^{-3.2\Delta month}$ and $P\propto e^{-3.3\Delta month}$ for (a) and (b), respectively, which is quite close to the one with all types of exit shown in Fig. \ref{fig1}e. }
\label{fig.appendix.exit}
\end{figure}

\begin{figure}[ht] \centering
\includegraphics[width=0.75\textwidth]{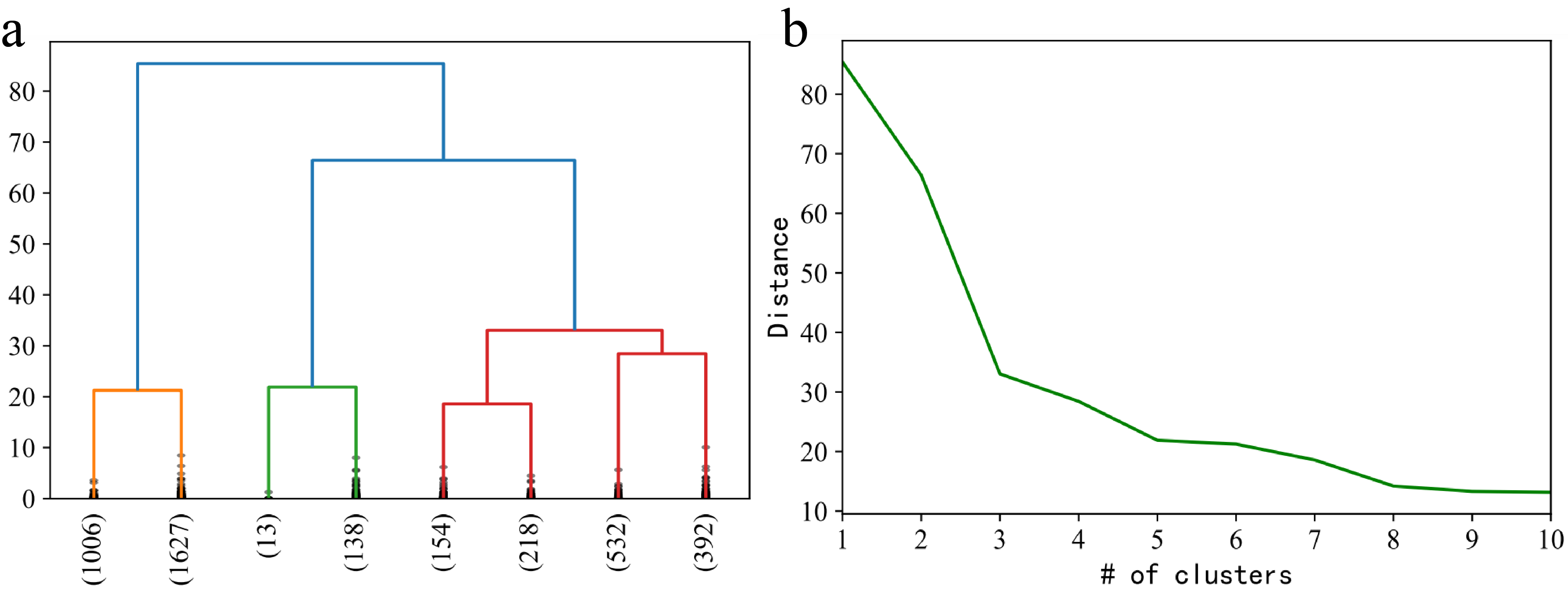}
\caption{(a) The hierarchical tree obtained by applying hierarchical clustering algorithm based on the evolution of k-shell values of all VC institutions. The digits on x-axis represent the size of certain branches. (b) The distance plot of determining the number of clusters, from which we can clearly see that 5 is the proper number of groups via the ``elbow'' point.} 
\label{fig.appendix.elbow}
\end{figure}

\begin{figure}[ht] \centering
\includegraphics[width=0.4\textwidth]{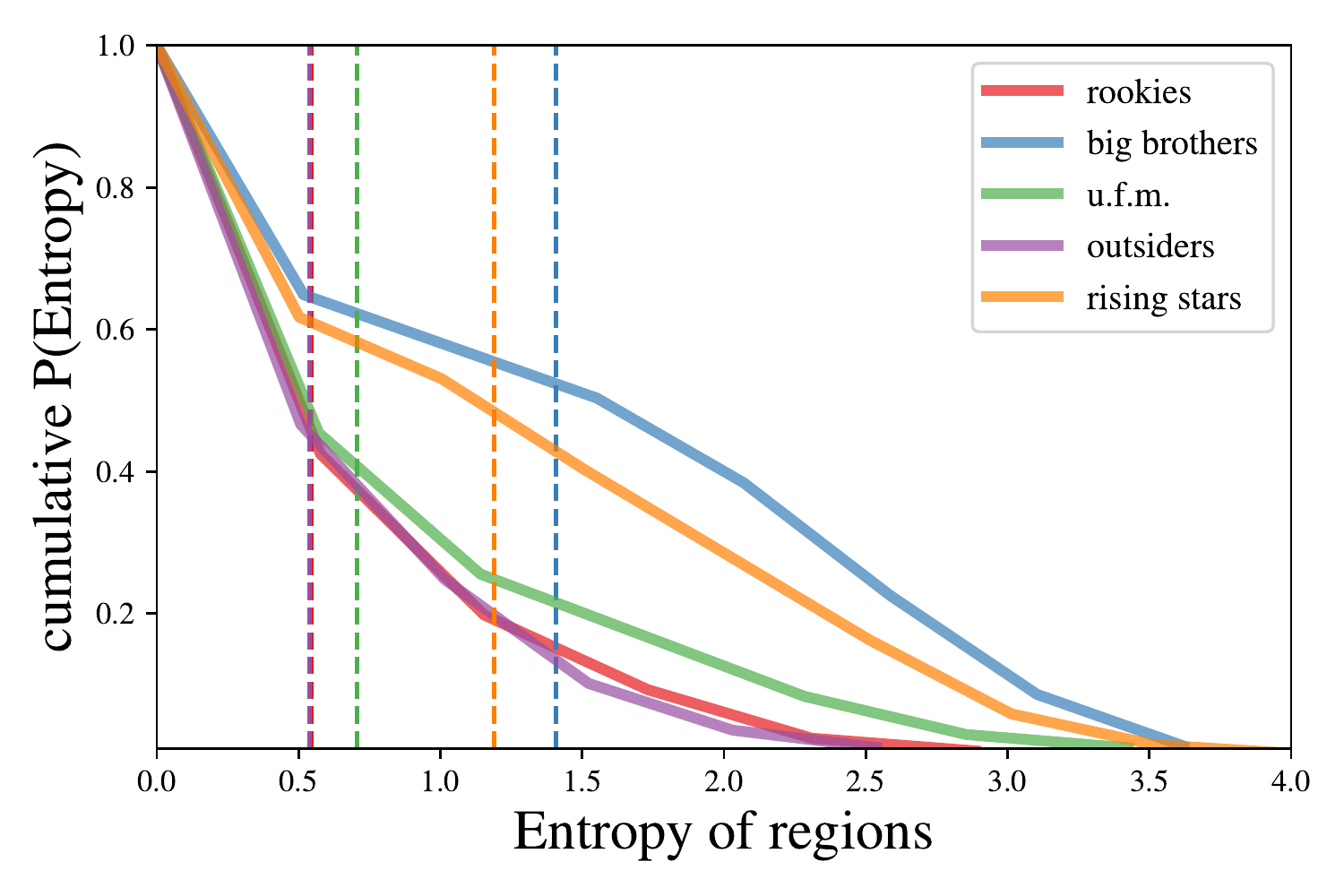}
\caption{
The distribution of region entropy of VC institutions in each group. Big brothers and rising stars invest in more regions and have a larger spatial diversity regarding their investments. Here, the region refers to the level of cities. 
} 
\label{fig.appendix.regionsdiversity}
\end{figure}


\section*{Acknowledgment.}
R.L. acknowledges helpful discussions with Prof. Haifeng Du from Xi'an Jiaotong University. 
This work is supported by NSFC under grant No. 61903020. L.Z. acknowledges support by National Natural Science Foundation of China under grant No. 71901171, Startup funding of NWPU G2021KY05101, ``the Fundamental Research Funds for the Central Universities'' 3102021XJS01, Shaanxi Provincial Soft Science Project 2022KRM111, and Shaanxi Provincial Social Science Foundation 2022R016. C.Z. acknowledges support by the Natural Science Foundation of Hebei (No. F2020205012), the Youth Top Talent Project of Hebei Education Department (No. BJ2020035), and Science Foundation of Hebei Normal University (No. L2023K04).. 

\section*{Authors' contributions.} R.L. conceived the study. R.L. and H.E.S. designed the study. C.C. and J.L. acquired the data. J.L. and R.L. conducted experiments and analyzed the results. R.L. was the leading writer of the manuscript, and all the authors reviewed the manuscript. 

\section*{Competing interests.} The authors declare that they have no competing interests.

\end{document}